\documentstyle[aps, epsfig, axodraw]{revtex}

\def\xslash#1{{\rlap{$#1$}/}}
\def\half{\frac{1}{2}}
\def\beq{\begin{equation}}
\def\eeq{\end{equation}}
\def\beqa{\begin{eqnarray}}
\def\eeqa{\end{eqnarray}}
\def\iar{\begin{array}{l}}
\def\ear{\end{array}}

\begin{document}
\draft
\title{A New Renormalization Scheme of Fermion Field in Standard Model}
\author{Yong Zhou$^{a}$}
\address{$^a$ Institute of Theoretical Physics, Chinese Academy of Sciences, 
         P.O. Box 2735, Beijing 100080, China}
\date{Jan 10, 2003}
\maketitle

\begin{abstract}
In order to obtain proper wave-function renormalization constants for unstable
fermion and consist with Breit-Wigner formula in the resonant region, We have 
assumed an extension of the LSZ reduction formula for unstable fermion and adopted
on-shell mass renormalization scheme in the framework of without field 
renormaization. The comparison of gauge dependence of physical amplitude between 
on-shell mass renormalization and complex-pole mass renormalization has been 
discussed. After obtaining the fermion wave-function renormalization constants, we
extend them to two matrices in order to include the contributions of off-diagonal 
two-point diagrams at fermion external legs for the convenience
of calculations of S-matrix elements.
\end{abstract}


\section{Introduction}

We are interested in the LSZ reduction formula \cite{c1} in the case of unstable
particle, for the purpose of obtaining a proper wave function renormalization
constant(wrc.) for unstable fermion. In this process we have demanded the form of the 
unstable particle's propagator in the resonant region resembles the relativistic 
Breit-Wigner formula. Therefore the on-shell mass 
renormalization scheme is a suitable choice compared with the complex-pole mass
renormalization scheme\cite{c2}, as we will discuss in section 2. 
We also discard the field 
renormalization constants which have been proven at question in the case of
having fermions mixing\cite{c3}. The 
layout of this article is as follows. First we discuss the problem arising from the
introduction of field renormalization constants in the case of
having fermions mixing. Next we introduce the extended
LSZ reduction formula for unstable fermion and obtain the proper fermion wrc..
Next we discuss the difference between different renormalization schemes
and the gauge dependence of physical amplitude. Next we change the
fermion wrc. into two matrices for the purpose of including the corrections of 
fermion external legs, thus simplify the calculations of S-matrix elements. Finally
we give our conclusion.

Along with the improvement of the precision of experiment, the radiative corrections
need to be calculated more precisely, especially for the imaginary part of Feynman
diagrams. For unstable particle the imaginary part of their self-energy 
functions should be considered in their wrc.\cite{c3}. Firstly let us introduce 
some notation in order to write them down. The bare fermion field is renormalized 
as follows

\beq 
  \Psi_{0} = Z^{\half}\Psi \,, \hspace{4mm} 
  \bar{\Psi}_{0} = \bar{\Psi}\bar{Z}^{\half}\,,
\eeq
with the field renormalization constants $Z^{\half}$ and $\bar{Z}^{\half}$\cite{c3}

\beq 
  Z^{\half} = Z^{L\half}\gamma_L + Z^{R\half}\gamma_R\,,\hspace{4mm}
  \bar{Z}^{\half} = \bar{Z}^{L\half}\gamma_{R} + \bar{Z}^{R\half}\gamma_{L}\,.  
\eeq
with $\gamma_L$ and $\gamma_R$ the left- and right-handed projection operator.
Within one-loop accuracy we have $Z^{\half} = 1 + \half\delta Z$ and
$\bar{Z}^{\half} = 1 + \half\delta\bar{Z}$. Thus the inverse fermion propagator is 
as follows (the letters $i,j$ denote the family indices)

\beq 
  -iS_{ij}^{-1}(\xslash p) = (\xslash p - m_{i}) \delta_{ij} + \hat{\Sigma}_{ij}
  (\xslash p) \,, 
\eeq
with the one-loop renormalized self-energy

\beqa
  \begin{array}{c}
  \hat{\Sigma}_{ij}(p) = \xslash p (\Sigma^L_{ij}(p^2)\gamma_{L} + \Sigma^R_{ij}(p^2)
  \gamma_{R}) + \Sigma^S_{ij}(p^2)(m_i \gamma_{L} + m_j \gamma_{R}) + 
  \\ \hspace{-2mm} 
  \half\delta \bar{Z}_{ij}(\xslash p - m_{j}) + \half(\xslash p - m_{i})
  \delta Z_{ij} - \delta m_{i}\delta _{ij}\,.  
  \ear  
\eeqa

If we introduce the field renormalization conditions that the fermion two-point 
functions are diagonal when the external-line momentums are on their mass shell, 

\beqa
  \left.
  \begin{array}{r}
  S_{ij}^{-1}(p)u_j (p)| _{p^2 = m_j^2} = 0 \\
  \bar{u}_{i}(p^{\prime})S_{ij}^{-1}(p^{\prime})| _{p^{\prime 2} = m_i^2} = 0 
  \ear 
  \right\} \, i\, \not= \, j\,.
\eeqa
with $\bar{u}_i$ and $u_j$ the Dirac spinors, we will encounter a bad thing 
that there is no solution for the above equations if we keep the "pseudo-hermiticity" 
relationship between $\bar{Z}^{\half}$ and $Z^{\half}$\cite{c3}

\beq
  \bar{Z}^{\half} = \gamma^{0} Z^{\half\dagger} \gamma^{0} 
\eeq
This is because of the existence of the imaginary part of the fermion two-point 
functions,
which makes $\Sigma^{L}_{ij}$, $\Sigma^{R}_{ij}$ and $\Sigma^{S}_{ij}$ un-Hermitian
\cite{c4}. As we know the relationship of Eq.(6) comes from the hermiticity of 
Lagrangian, which guarantees the S-matrix is unitary, thus cannot be broken. 

A method to solve this problem is to introduce two sets of field renormalization
constants, one set for the external-line particles, the other set for the inner-line 
particles in a Feynman diagram\cite{c3}. But it brings some complexity and cannot 
explain why we couldn't find a consistent field renormalization constant for all
particles. 

\section{Wave-function Renormalization Constants}

Another method to solve the above problem is to abandon field renormalization,
since particle fields don't belong to physical parameters thus don't need to be
renormalized. One effect of this method is that the calculations of Feynman 
diagrams becomes simpler than before. The other effect will be seen in section 3. 

Since there aren't field renormalization constants, we need to find a way to 
determine the fermion wrc. 
We demand the propagator of unstable particle has the following form when the 
momentum is on mass shell

\vspace{5mm} 
\beq 
  \begin{picture}(48,14)
    \ArrowLine(0,5)(16,5)
    \GCirc(23,5){7}{0.5}
    \ArrowLine(30,5)(48,5)
  \end{picture} \, \sim \, \frac{1}{p^2 - m^2 + i m \Gamma (p^2)}
\eeq
\vspace{3mm}

\hspace{-3mm}where $m\Gamma(p^2)$ is a real number which 
represents the imaginary part of the self-energy function times 
the particle's wrc.. This form resembles the relativistic Breit-wigner formula,
thus will be a reasonable assumptions. Under this assumption only the on-shell
mass renormalization scheme is suitable, the other scheme, complex-pole 
mass renormalization scheme, cannot satisfy Eq.(7). We can see this point 
in the boson case.

The mass definition in complex-pole scheme is based on the complex-valued position of 
the propagator's pole\cite{c2}:

\beq
  \bar{s} \,=\, M_0^2 + A(\bar{s}) \,.
\eeq
with $A(p^2)$ the self-energy function of boson. Writing 
$\bar{s}=m_2^2 - i m_2 \Gamma_2$, the mass and width of the unstable particle may be 
identified with $m_2$ and $\Gamma_2$, respectively. Given $m_2$ and $\Gamma_2$, 
there are two other definitions:

\begin{eqnarray*} \iar
  m_1 \,=\, \sqrt{m_2^2 + \Gamma_2^2}\,, \Gamma_1 \,=\, \frac{m_1}{m_2}\Gamma_2 \,, 
  \\ \bar{s} \hspace{3mm} =\, (m_3 - i \frac{\Gamma_3}{2})^2 \,.
\ear \end{eqnarray*}
The form of the propagator in resonant region is

$$ \frac{1}{p^2-M_0^2-A(p^2)}\,\sim\,\frac{1}{m^2-M_0^2-A(m^2)+(p^2-m^2)
   (1-A^{\prime}(m^2))}\,=\,\frac{(1-A^{\prime}(m^2))^{-1}}{p^2-m^2+B} $$
with $B=(m^2-M_0^2-A(m^2))/(1-A^{\prime}(m^2))$. Through simple calculations we find
all of these three definitions of mass cannot make $B$ a pure imaginary number. So 
the complex-pole mass renormalization scheme isn't suitable for the assumption of 
Eq.(7). 

Because the unstable fermion's propagator near resonant region must contain
Dirac matrices, Eq.(7) isn't enough. We further assume it has the following form:

\beq
  \frac{i}{\xslash p - m_i + \hat{\Sigma}_{ii}(\xslash p)}\, 
  \begin{array}{c} \vspace{-7mm} \\ p^2 \rightarrow m_i^2 \\ \vspace{-6mm} \\ 
  -\hspace{-2mm}-\hspace{-2mm}-\hspace{-2mm}-\hspace{-2mm}
  -\hspace{-3mm}\longrightarrow \ear \,
  \frac{i Z_i^{\half} (\xslash p + m_i + i x)\bar{Z}_i^{\half}} 
  {p^2 - m_i^2 + i m_i \Gamma_i (p^2)}\,.
\eeq
where $Z_i^{\half}$ and $\bar{Z}_i^{\half}$ are the fermion wrc., $x$ is a quantity 
which has
nothing to do with our present calculations. The LSZ reduction formula which relates
the Green functions to S-matrix elements may have such form:

\begin{eqnarray} \iar
  \hspace{17mm}
  \int dx_1^4 dx_2^4 e^{i p_1 x_1} e^{i p_2 x_2} <\Omega |T\bar{\psi}(x_1)\psi(x_2)
  \cdot\cdot\cdot | \Omega>\, \\ \vspace{1mm} \\ 
  \begin{array}{c} \vspace{-7mm} \\ p_i^2 \rightarrow m_i^2 \\ \vspace{-6mm} \\ 
  -\hspace{-2mm}-\hspace{-2mm}-\hspace{-2mm}-\hspace{-2mm}
  -\hspace{-3mm}\longrightarrow \ear \,
  \frac{Z_1^{\half} (\xslash p_1 + m_1 + i x_1)}{p_1^2 - m_1^2 + i m_1 \Gamma_1 
  (p_1^2)}<\bar{\psi}(p_1)\cdot\cdot\cdot |S| \cdot\cdot\cdot\psi(p_2)>
  \frac{(\xslash p_2 + m_2 + i x_2)\bar{Z}_2^{\half}}
  {p_2^2 - m_2^2 + i m_2 \Gamma_2 (p_2^2)}\, .
\ear \end{eqnarray}
here we have assumed the Heisenberg field $\bar{\psi}(x_1)$ can be transformed into
the out-state $<\hspace{-1mm}\bar{\psi}(p_1)|$ with an additional coefficient 
$\frac{Z_1^{\half} (\xslash p_1+m_1+i x_1)}{p_1^2-m_1^2+i m_1 \Gamma_1 (p_1^2)}$, and
the Heisenberg field $\psi(x_2)$ can be transformed into the in-state 
$|\psi(p_2)\hspace{-1mm}>$ with an additional coefficient 
$\frac{(\xslash p_2+m_2+i x_2)\bar{Z}_2^{\half}}{p_2^2-m_2^2+i m_2\Gamma_2(p_2^2)}$. 
According to Eq.(9), we can sum up the corrections to each external legs in a Green 
function and analyse the Green function into the amputated part times the 
corresponding propagators. That is to say

\beqa \iar
  \hspace{17mm}
  \int dx_1^4 dx_2^4 e^{i p_1 x_1} e^{i p_2 x_2} <\Omega |T\bar{\psi}(x_1)\psi(x_2)
  \cdot\cdot\cdot | \Omega>\, \\ \vspace{1mm} \\
  \begin{array}{c} \vspace{-7mm} \\ p_i^2 \rightarrow m_i^2 \\ \vspace{-6mm} \\ 
  -\hspace{-2mm}-\hspace{-2mm}-\hspace{-2mm}-\hspace{-2mm}
  -\hspace{-3mm}\longrightarrow \ear \,
  \frac{Z_1^{\half} (\xslash p_1 + m_1 + i x_1)\bar{Z}_1^{\half}} 
  {p_1^2 - m_1^2 + i m_1 \Gamma_1 (p_1^2)}
  <\Omega |T\bar{\psi}(p_1)\psi(p_2) \cdot\cdot\cdot | \Omega >_{amp}
  \frac{Z_2^{\half} (\xslash p_2 + m_2 + i x_2)\bar{Z}_2^{\half}} 
  {p_2^2 - m_2^2 + i m_2 \Gamma_2 (p_2^2)}\, .
\ear \eeqa
where the subscript $amp$ refers to the amputated Green function. Thus the LSZ 
reduction formula with two fermions at external legs will be:

\beqa
  < p_1 ... p_n | S | k_1 ... k_m > \,=\, Z^{\frac{n+m-2}{2}} 
  \bar{u}(p_1)\bar{Z}_f^{\half} \hspace{1mm}
  \begin{picture}(50,30)
    \BCirc(25,4){12}
    \Text(25,4)[]{\small Amp.}
    \ArrowLine(5,-20)(18,-7)
    \ArrowLine(18,15)(5,28)
    \ArrowLine(45,-20)(32,-7)
    \ArrowLine(32,15)(45,28)
    \Text(25,-18)[]{$\cdot \cdot \cdot$}
    \Text(25,26)[]{$\cdot \cdot \cdot$}
    \Text(0,-20)[]{$k_1$}
    \Text(0,28)[]{$p_1$}
    \Text(52,-20)[]{$k_m$}
    \Text(52,28)[]{$p_n$}
  \end{picture} \hspace{1mm} 
  Z_f^{\half} u(k_1) \, .
\eeqa
\vspace{4mm}

\hspace{-4mm}with $\bar{Z}_f^{\half}$ and $Z_f^{\half}$ the fermion wrc. and $Z$ the
boson wrc.. The fermion spinors $\bar{u}(p_1)$ and $u(k_1)$ have been added
to the right hand side of the above equation, as are demanded.

In actual calculations, the fermion propagator needs to be transformed into the form 
of Eq.(9). To start with, we introduce the renormalized fermion self-energy functions 
as follows
  
\beq
  \hat{\Sigma}_{ii}\,=\, \xslash p (\Sigma^L_{ii}(p^2)\gamma_L + \Sigma^R_{ii}(p^2)
  \gamma_R) + \Sigma^{S,L}_{ii}(p^2)\gamma_L + \Sigma^{S,R}_{ii}(p^2)\gamma_R - 
  \delta m_i \,.
\eeq
Although the self-energy functions contain only the contributions of 
one-particle-irreducible (1PI) diagrams in usual sense, it isn't denied in principle 
to include
the contributions of reducible diagrams in the self-energy functions. In the present
scheme the non-diagonal two-point functions aren't zero when external
momentum is on shell, since there aren't the conditions of Eqs.(5). Thus the diagonal
self-energy functions will contain the contributions of the reducible diagrams:

\vspace{5mm} 
\beq
  \begin{picture}(420,10)
    \ArrowLine(25,5)(38,5)
    \GCirc(42,5){4}{0.5}
    \ArrowLine(46,5)(59,5)
    \Text(30,13)[]{$i$}
    \Text(54,13)[]{$i$}
    \Text(67,5)[]{$+$}
    \Text(80,1)[]{$\stackrel{\sum}{k}$}
    \ArrowLine(85,5)(98,5)
    \GCirc(102,5){4}{0.5}
    \ArrowLine(106,5)(119,5)
    \GCirc(123,5){4}{0.5}
    \ArrowLine(127,5)(140,5)
    \Text(90,13)[]{$i$}
    \Text(112,13)[]{$k$}
    \Text(135,13)[]{$i$}
    \Text(148,5)[]{$+$}
    \Text(161,1)[]{$\stackrel{\sum}{k,l}$}
    \ArrowLine(166,5)(179,5)
    \GCirc(183,5){4}{0.5}
    \ArrowLine(187,5)(200,5)
    \GCirc(204,5){4}{0.5}
    \ArrowLine(208,5)(221,5)
    \GCirc(225,5){4}{0.5}
    \ArrowLine(229,5)(242,5)
    \Text(171,13)[]{$i$}
    \Text(193,13)[]{$k$}
    \Text(214,13)[]{$l$}
    \Text(237,13)[]{$i$}
    \Text(250,5)[]{$+$}
    \Text(266,5)[]{$\cdot \cdot \cdot$}
    \Text(282,5)[]{$+$}
    \Text(298,1)[]{$\stackrel{\sum}{k,l,\cdot \cdot \cdot,m}$}
    \ArrowLine(306,5)(319,5)
    \GCirc(323,5){4}{0.5}
    \ArrowLine(327,5)(340,5)
    \GCirc(344,5){4}{0.5}
    \ArrowLine(348,5)(361,5)
    \Text(371,5)[]{$\cdot \cdot \cdot$}
    \ArrowLine(377,5)(390,5)
    \GCirc(395,5){4}{0.5}
    \ArrowLine(399,5)(412,5)
    \Text(311,13)[]{$i$}
    \Text(333,13)[]{$k$}
    \Text(354,13)[]{$l$}
    \Text(383,13)[]{$m$}
    \Text(407,13)[]{$i$}
    \Text(380,-15)[]{$k,l,\cdot \cdot \cdot,m \not= i$}
  \end{picture} 
\eeq
\vspace{3mm}

\hspace{-4mm}where the disc is the 1PI diagram. The condition 
$k,l,\cdot \cdot \cdot,m \not= i$
avoids the reduplication in actual calculations. Now we write the inverse fermion 
propagator as follows

\beq 
  -iS_{ii}^{-1}(\xslash p) = \xslash p - m_{i} + \hat{\Sigma}_{ii}(\xslash p)
  = \xslash p (a\gamma_L+b\gamma_R) + c\gamma_L + d\gamma_R
\eeq
with

\beqa \iar
  a\,=\,1+\Sigma^L_{ii}(p^2)\,,\hspace{16mm}b\,=\,1+\Sigma^R_{ii}(p^2)\,,\\
  c\,=\,\Sigma^{S,L}_{ii}(p^2)-m_i-\delta m_i\,,\,
  d\,=\,\Sigma^{S,R}_{ii}(p^2)-m_i-\delta m_i\,.
\ear \eeqa
and therefore after some algebra\cite{c3}

\beq 
  -iS_{ii}^{-1}(\xslash p) = \frac{\xslash p (a\gamma_L+b\gamma_R) - d\gamma_L
  - c\gamma_R}{p^2 a b - c d}
\eeq
That's to say

\beq
  \frac{i}{\xslash p - m_i + \hat{\Sigma}_{ii}(\xslash p)}\, = \,
  \frac{i(m_i + \delta m_i + \xslash p \gamma_L (1 + \Sigma^L_{ii}) + \xslash p 
  \gamma_R
  (1 + \Sigma^R_{ii}) - \gamma_L \Sigma^{S,L}_{ii} - \gamma_R \Sigma^{S,R}_{ii})}
  {p^2(1 + \Sigma^L_{ii})(1 + \Sigma^R_{ii}) - (m_i + \delta m_i - \Sigma^{S,L}_{ii})
  (m_i + \delta m_i - \Sigma^{S,R}_{ii})} \,.
\eeq
Adopting the on-shell mass renormalization scheme\footnote{There are many articles
discussing the gauge dependence of the on-shell mass renormalization 
scheme\cite{c5}, we will discuss this problem in section 4}, the mass 
renormalization condition is 

\beq
  Re[m_i^2 (1+\Sigma^L_{ii}(m_i^2))(1+\Sigma^R_{ii}(m_i^2))-(m_i+\delta m_i- 
  \Sigma^{S,L}_{ii}(m_i^2))(m_i + \delta m_i - \Sigma^{S,R}_{ii}(m_i^2)) ]\,=\, 0
\eeq
Expanding the denominator of Eq.(18) at $p^2=m_i^2$ according to Eq.(9), we obtain

\beqa
  \frac{i}{\xslash p - m_i + \hat{\Sigma}_{ii}(\xslash p)}   
  \begin{array}{c} \vspace{-7mm} \\ p^2 \rightarrow m_i^2 \\ \vspace{-6mm} \\ 
  -\hspace{-2mm}-\hspace{-2mm}-\hspace{-2mm}-\hspace{-2mm}
  -\hspace{-3mm}\longrightarrow \ear \,
  \frac{i(m_i + \delta m_i + \xslash p \gamma_L 
  (1 + \Sigma^L_{ii}(m_i^2)) + \xslash p \gamma_R (1 + \Sigma^R_{ii}(m_i^2)) - 
  \gamma_L \Sigma^{S,L}_{ii}(m_i^2) - \gamma_R \Sigma^{S,R}_{ii}(m_i^2))}
  {(p^2 -m_i^2 + i m_i \Gamma_i(p^2))A}
\eeqa
\vspace{1mm} 

\hspace{-4mm}with 

\beqa \iar
  A\,=\,Re[1 + \Sigma^L_{ii} + \Sigma^R_{ii} + \Sigma^L_{ii}\Sigma^R_{ii} + 
  m_i^2 \frac{\partial}{\partial p^2}(\Sigma^L_{ii} + \Sigma^R_{ii} + 
  \Sigma^L_{ii}\Sigma^R_{ii}) + \\ \hspace{8mm} 
  \frac{\partial \Sigma^{S,L}_{ii}}{\partial p^2}(m_i+\delta m_i-\Sigma^{S,R}_{ii})+
  \frac{\partial \Sigma^{S,R}_{ii}}{\partial p^2}(m_i+\delta m_i-\Sigma^{S,L}_{ii})]
\ear \eeqa

\beq
  m_i \Gamma_i(p^2)=\frac{1}{A} Im[p^2(\Sigma^L_{ii}(p^2)+\Sigma^R_{ii}(p^2)+
  \Sigma^L_{ii}(p^2)\Sigma^R_{ii}(p^2))+(m_i+\delta m_i)(\Sigma^{S,L}_{ii}(p^2)+
  \Sigma^{S,R}_{ii}(p^2))-\Sigma^{S,L}_{ii}(p^2)\Sigma^{S,R}_{ii}(p^2)]
\eeq
Noted we have discarded the regular terms in Eq.(20). 

From Eq.(9) and (20) we obtain

\beqa \iar
  Z_i^{L\half}\bar{Z}_i^{L\half}\,=\, (1 + \Sigma^R_{ii}(m_i^2))/A \,, \hspace{33mm}
  Z_i^{R\half}\bar{Z}_i^{R\half}\,=\, (1 + \Sigma^L_{ii}(m_i^2))/A \,, \\ 
  Z_i^{L\half}\bar{Z}_i^{R\half}\,=\, (m_i+\delta m_i-\Sigma^{S,R}_{ii}(m_i^2))/
  (A(m_i + i x)) \,, Z_i^{R\half}\bar{Z}_i^{L\half}\,=\, (m_i+\delta m_i- 
  \Sigma^{S,L}_{ii}(m_i^2))/(A(m_i + i x))\,.
\ear \eeqa
\hspace{-1mm}In order to solve these equations, we firstly obtain a equation 
about $x$

\beq
  (m_i + i x)^2 (1 + \Sigma^L_{ii}(m_i^2))(1 + \Sigma^R_{ii}(m_i^2))\,=\, 
  (m_i + \delta m_i -   \Sigma^{S,L}_{ii}(m_i^2))(m_i + \delta m_i - 
  \Sigma^{S,R}_{ii}(m_i^2))\,.
\eeq
Replacing $m_i + i x$ in Eqs.(23) by this relationship, we obtain

{\large \beqa \iar
  {\tiny Z}_i^{L\half}\bar{{\tiny Z}}_i^{L\half}\,=\, \frac{1 + 
  \Sigma^R_{ii}(m_i^2)}{A}\,,
  {\tiny Z}_i^{R\half}\bar{{\tiny Z}}_i^{R\half}\,=\, \frac{1 + 
  \Sigma^L_{ii}(m_i^2)}{A}\,, \\
  \frac{Z_i^L}{Z_i^R}\,=\, \frac{(m_i + \delta m_i - \Sigma^{S,R}_{ii}(m_i^2))
  (1 + \Sigma^R_{ii}(m_i^2))} {(m_i + \delta m_i - \Sigma^{S,L}_{ii}(m_i^2))
  (1 + \Sigma^L_{ii}(m_i^2))}\,, \\
  \frac{\bar{Z}_i^L}{\bar{Z}_i^R}\,=\, \frac{(m_i + \delta m_i - \Sigma^{S,L}_{ii}
  (m_i^2))
  (1 + \Sigma^R_{ii}(m_i^2))} {(m_i + \delta m_i - \Sigma^{S,R}_{ii}(m_i^2))
  (1 + \Sigma^L_{ii}(m_i^2))}\,.
\ear \eeqa }

Because we haven't introduced the fermion field renormalization constants, which 
contain Dirac matrices thus may change the Lorentz structure of the fermion's 
self-energy functions, the self-energy functions in our scheme will keep the Lorentz
invariant structure:

\beq
  \hat{\Sigma}_{ij}(\xslash p) = \xslash p (\Sigma^L_{ij}(p^2)\gamma_{L} + 
  \Sigma^R_{ij}(p^2)
  \gamma_{R}) + \Sigma^S_{ij}(p^2)(m_{0,i} \gamma_{L} + m_{0,j} \gamma_{R}) - 
  \delta m_i \delta_{ij}\,.
\eeq
where $m_{0,i}$ and $m_{0,j}$ the bare fermion masses. It is obvious that

\beq
  \Sigma^{S,R}_{ii}\,=\,\Sigma^{S,L}_{ii}\,\equiv\, m_i \Sigma_{ii}^S
\eeq 
Therefore Eqs.(25) can be simplified as follows

\beqa \iar
  Z_i^{L\half}\bar{Z}_i^{L\half}\,=\, (1 + \Sigma^R_{ii}(m_i^2))/A \, \hspace{20mm}
  Z_i^{R\half}\bar{Z}_i^{R\half}\,=\, (1 + \Sigma^L_{ii}(m_i^2))/A \,, \\
  Z_i^L/Z_i^R\hspace{3mm}=\, (1+\Sigma^R_{ii}(m_i^2))/(1+\Sigma^L_{ii}(m_i^2))\,, 
  \bar{Z}_i^L/ \bar{Z}_i^R\hspace{3mm}=\,(1+\Sigma^R_{ii}(m_i^2))/
  (1+\Sigma^L_{ii}(m_i^2))\,.
\ear \eeqa 
\hspace{-1mm}Because the four equations aren't independent, we need to find an 
additional condition. From the strong symmetry between $\bar{Z}$ and $Z$ manifested 
in the above equations, a reasonable assumption is to set 
$\bar{Z}_i^L\,=\, Z_i^L\,, \bar{Z}_i^R\,=\, Z_i^R$, which is also the case of stable 
fermion. Therefore the final results are

\beqa \iar
  \bar{Z}_i^L \,=\, Z_i^L \,=\, (1 + \Sigma^R_{ii}(m_i^2))/A \,, \\
  \bar{Z}_i^R \,=\, Z_i^R \,=\, (1 + \Sigma^L_{ii}(m_i^2))/A \,.
\ear \eeqa
and

\beq
 \frac{\delta m_i}{m_i}\,=\, \sqrt{Re[(1+\Sigma^L_{ii}(m_i^2))
 (1+\Sigma^R_{ii}(m_i^2))]+Im[\Sigma^S_{ii}(m_i^2)]^2}+Re[\Sigma^S_{ii}(m_i^2)]-1
\eeq

It is very easy to give the explicit one-loop level results. The wrc., mass
counterterm and decay rate are listed below:

\beqa \iar
  \bar{Z}^L_i=Z^L_i=1-Re[\Sigma^L_{ii}(m_i^2)+m_i^2\frac{\partial}{\partial p^2}
  (\Sigma^L_{ii}(p^2)+\Sigma^R_{ii}(p^2)+2\Sigma^S_{ii}(p^2))|_{p^2=m_i^2}]+
  i Im[\Sigma^R_{ii}(m_i^2)]\,, \\
  \bar{Z}^R_i=Z^R_i=1-Re[\Sigma^R_{ii}(m_i^2)+m_i^2\frac{\partial}{\partial p^2}
  (\Sigma^L_{ii}(p^2)+\Sigma^R_{ii}(p^2)+2\Sigma^S_{ii}(p^2))|_{p^2=m_i^2}]+
  i Im[\Sigma^L_{ii}(m_i^2)]\,.
\ear \eeqa

\beq
  \hspace{1mm}
  \delta m_i\,=\, \frac{m_i}{2}Re[\Sigma^L_{ii}(m_i^2)+\Sigma^R_{ii}(m_i^2)+
  2\Sigma^S_{ii}(m_i^2)]
\eeq 

\beq
  m_i\Gamma_i\,=\, m_i^2 Im[\Sigma^L_{ii}(m_i^2)+\Sigma^R_{ii}(m_i^2)+
  2\Sigma^S_{ii}(m_i^2)]
\eeq
Another meaningful result concerning $x$ (which appears in Eq.(9)) is:

\beq
  x\,=\, -\Gamma_i/2
\eeq

Until now we haven't discussed the Optical Theorem. It is a very complex problem, so 
we want to leave it for the future work. Here we only point out that 
Eq.(33) is really the fermion decay rate at one-loop level\cite{c3}. 

\section{Comparison between Different Renormalization Schemes}

Adopting which mass renormalization scheme and whether introducing field 
renormalization constants determine the difference of the renormalization schemes. 
Here we will list three other schemes and compare their results.

Firstly we consider the renormalization scheme which adopts the complex-pole mass
renormalization scheme and doesn't introduce field renormalization constants. In 
this case Eq.(9) isn't satisfied, but we can assume $\Gamma_i$ is a complex number
and independent of the propagator's momentum. Thus all of the results associated 
with fermion wrc. in previous section keep unchanged, except for removing the 
operator $Re$ from $A$, because in this case it is no sense to treat the
real part and the imaginary part of the denominator of the fermion propagator 
differently. Therefore we obtain

\beqa \iar
  \bar{Z}_i^L \,=\, Z_i^L \,=\, (1 + \Sigma^R_{ii}(m_i^2))/A_1  \\
  \bar{Z}_i^R \,=\, Z_i^R \,=\, (1 + \Sigma^L_{ii}(m_i^2))/A_1 
\ear \eeqa
with

\beq
  A_1=1 + \Sigma^L_{ii} + \Sigma^R_{ii} + \Sigma^L_{ii}\Sigma^R_{ii} + 
  m_i^2 \frac{\partial}{\partial p^2}(\Sigma^L_{ii} + \Sigma^R_{ii} + 
  \Sigma^L_{ii}\Sigma^R_{ii}) +  
  \frac{\partial \Sigma^{S,L}_{ii}}{\partial p^2}(m_i+\delta m_i-\Sigma^{S,R}_{ii})+
  \frac{\partial \Sigma^{S,R}_{ii}}{\partial p^2}(m_i+\delta m_i-\Sigma^{S,L}_{ii})
\eeq
On the other hand, $\delta m_i$ and $\Gamma_i$ are different from the previous 
results very much. But at one-loop level they are the same as Eq.(32) and (33).

Next we study the renormalization scheme which adopts the complex-pole mass
renormalization scheme and has introduced field renormalization constants. 
We still assume $\Gamma_i$ is a complex number and independent of the propagator's 
momentum. Because the field renormalization constants have been introduced, the 
fermion self-energy functions will contain their contributions. We can write 
down the fermion self-energy functions in such form:

\beqa \iar
  \begin{picture}(52,14)
    \ArrowLine(0,5)(18,5)
    \GCirc(26,5){8}{0.5}
    \ArrowLine(34,5)(52,5)
    \Text(18,5)[]{$\times$}
    \Text(34,5)[]{$\times$}
    \Text(6,13)[]{$i$}
    \Text(46,13)[]{$i$}
  \end{picture} \,=\,
  \Sigma^L_{ii} \xslash p \gamma_L + \Sigma^R_{ii} \xslash p \gamma_R + 
  \Sigma^{S,L}_{ii} \gamma_L + \Sigma^{S,R}_{ii} \gamma_R \,, \\
  \hspace{20mm} \,\equiv \, 
  \bar{Z}_{ii}^{\half} (\Sigma^{\prime L}_{ii} \xslash p \gamma_L + 
  \Sigma^{\prime R}_{ii} \xslash p \gamma_R + \Sigma^{\prime S,L}_{ii} \gamma_L +
  \Sigma^{\prime S,R}_{ii} \gamma_R) Z_{ii}^{\half}\,.
\ear \eeqa
where $\times$ denotes the field renormalization constants. This form can be easily
understood if we regard it as an application of the LSZ reduction formula Eq.(12) at
two-point Green functions. The new self-energy function $\Sigma^{\prime}$ is 
called "pure" self-energy function by us. It is easy to obtain from the above 
equation that

\beqa \iar
  \Sigma^L_{ii}\hspace{3mm}=\,\bar{Z}_{ii}^{L\half}Z_{ii}^{L\half}
  \Sigma^{\prime L}_{ii}\,, \hspace{3mm}
  \Sigma^R_{ii}\hspace{3mm}=\,\bar{Z}_{ii}^{R\half}Z_{ii}^{R\half}
  \Sigma^{\prime R}_{ii}\,, \\
  \Sigma^{S,L}_{ii}\,=\,\bar{Z}_{ii}^{R\half}Z_{ii}^{L\half}
  \Sigma^{\prime S,L}_{ii}\,, 
  \Sigma^{S,R}_{ii}\,=\,\bar{Z}_{ii}^{L\half}Z_{ii}^{R\half} 
  \Sigma^{\prime S,R}_{ii}\,.
\ear \eeqa
We can write down the inverse fermion propagator in the form of Eq.(15) with

\beqa \iar
  a \,=\, \bar{Z}_{ii}^{L\half} Z_{ii}^{L\half} + \Sigma_{ii}^L(p^2) \,, 
  \hspace{20mm}
  b \,=\, \bar{Z}_{ii}^{R\half} Z_{ii}^{R\half} + \Sigma_{ii}^R(p^2) \,, \\
  c \,=\, \Sigma_{ii}^{S,L}(p^2) - (m_i + \delta m_i) \bar{Z}_{ii}^{R\half} 
  Z_{ii}^{L\half}\,,
  d \,=\, \Sigma_{ii}^{S,R}(p^2) - (m_i + \delta m_i) \bar{Z}_{ii}^{L\half} 
  Z_{ii}^{R\half}\,.
\ear \eeqa
According to Eq.(17), the condition that in the limit $p^2\rightarrow m_i^2$ 
the chiral structure in the numerator must be canceled out leads to 

\beqa \begin{array}{c}
  \bar{Z}_{ii}^{L\half} Z_{ii}^{L\half} + \Sigma_{ii}^L(m_i^2)\,=\, 
  \bar{Z}_{ii}^{R\half} Z_{ii}^{R\half} + \Sigma_{ii}^R(m_i^2) \,, \\
  \Sigma_{ii}^{S,L}(m_i^2)-(m_i+\delta m_i)\bar{Z}_{ii}^{R\half}Z_{ii}^{L\half}\,=\,
  \Sigma_{ii}^{S,R}(m_i^2)-(m_i+\delta m_i)\bar{Z}_{ii}^{L\half}Z_{ii}^{R\half}\,.
\ear \eeqa
On the other hand the unit residue renormalization condition amounts to requiring

\beq
  a \,=\, a b + m_i^2 a \frac{\partial}{\partial p^2}b + m_i^2 b 
  \frac{\partial}{\partial p^2}a - c \frac{\partial}{\partial p^2}d -
  d \frac{\partial}{\partial p^2}c
\eeq
Using Eq.(39) and Eqs.(38), we have

\beqa \iar
  1=\bar{Z}_{ii}^{L\half}Z_{ii}^{L\half}[1+\Sigma_{ii}^{\prime L}+m_i^2
  (\frac{\partial}{\partial p^2}\Sigma_{ii}^{\prime L}+
  \frac{1+\Sigma_{ii}^{\prime L}}{1+\Sigma_{ii}^{\prime R}}
  \frac{\partial}{\partial p^2}\Sigma_{ii}^{\prime R})+
  \frac{m_i+\delta m_i-\Sigma_{ii}^{\prime S,R}}{1+\Sigma_{ii}^{\prime R}}
  \frac{\partial}{\partial p^2}\Sigma_{ii}^{\prime S,L}+
  \frac{m_i+\delta m_i-\Sigma_{ii}^{\prime S,L}}{1+\Sigma_{ii}^{\prime R}}
  \frac{\partial}{\partial p^2}\Sigma_{ii}^{\prime S,R}]
\ear \eeqa
Like the case in section 2 we need an additional condition to determine the field 
renormalization constants. If we adopt the same assumption as the ones in the 
previous section, 
$\bar{Z}_{ii}^{L}\,=\,Z_{ii}^L\,,\bar{Z}_{ii}^{R}\,=\,Z_{ii}^R$, we will obtain

\beqa \iar
  \bar{Z}_{ii}^L\,=\,Z_{ii}^L\,=\,(1+\Sigma^{\prime R}_{ii}(m_i^2))/A_2\,, \\
  \bar{Z}_{ii}^R\,=\,Z_{ii}^R\,=\,(1+\Sigma^{\prime L}_{ii}(m_i^2))/A_2\,. 
\ear \eeqa
with

\beqa \iar
  A_2\,=\,1+\Sigma^{\prime L}_{ii}+\Sigma^{\prime R}_{ii}+
  \Sigma^{\prime L}_{ii}\Sigma^{\prime R}_{ii}+m_i^2\frac{\partial}{\partial p^2}
  (\Sigma^{\prime L}_{ii}+\Sigma^{\prime R}_{ii}+
  \Sigma^{\prime L}_{ii}\Sigma^{\prime R}_{ii})+ \\ \hspace{10mm}
  \frac{\partial\Sigma^{\prime S,L}_{ii}}{\partial p^2}(m_i+\delta m_i-
  \Sigma^{\prime S,R}_{ii})+\frac{\partial\Sigma^{\prime S,R}_{ii}}{\partial p^2}
  (m_i+\delta m_i-\Sigma^{\prime S,L}_{ii})
\ear \eeqa
$A_2$ is the same as $A_1$, except for replacing the self-energy functions with
"pure" self-energy functions. But at one-loop level they are the same, so the 
one-loop results of Eqs.(43) are\cite{c3}

\beqa \iar
  \bar{Z}^L_i=Z^L_i=1-\Sigma^L_{ii}(m_i^2)-m_i^2\frac{\partial}{\partial p^2}
  (\Sigma^L_{ii}(p^2)+\Sigma^R_{ii}(p^2)+2\Sigma^S_{ii}(p^2))|_{p^2=m_i^2}\,, \\
  \bar{Z}^R_i=Z^R_i=1-\Sigma^R_{ii}(m_i^2)-m_i^2\frac{\partial}{\partial p^2}
  (\Sigma^L_{ii}(p^2)+\Sigma^R_{ii}(p^2)+2\Sigma^S_{ii}(p^2))|_{p^2=m_i^2}\,.
\ear \eeqa

Finally we consider the renormalization scheme which adopts the on-shell mass
renormalization scheme and has introduced field renormalization constants. In this 
scheme, the self-energy functions can still be written down in the form of Eqs.(37) 
and the renormalization conditions of Eqs.(40) keep unchanged. The only variation 
comes from the unit residue condition. In on-shell scheme we should treat the real
part and the imaginary part of the denominator of the fermion propagator in 
different ways. Thus the unit residue condition becomes

\beq
  a \,=\, Re[a b + m_i^2 a \frac{\partial}{\partial p^2}b + m_i^2 b 
  \frac{\partial}{\partial p^2}a - c \frac{\partial}{\partial p^2}d -
  d \frac{\partial}{\partial p^2}c]
\eeq
after some algebra we have

\beqa \iar
  1\,=\,\bar{Z}_{ii}^{L\half}Z_{ii}^{L\half}(1+\Sigma_{ii}^{\prime L})[1+Re[
  \frac{m_i^2}{1+\Sigma_{ii}^{\prime L}}\frac{\partial}{\partial p^2}
  \Sigma_{ii}^{\prime L}+\frac{m_i^2}{1+\Sigma_{ii}^{\prime R}}
  \frac{\partial}{\partial p^2}\Sigma_{ii}^{\prime R}+ \\ \hspace{8mm}
  \frac{m_i+\delta m_i-\Sigma_{ii}^{\prime S,R}}
  {(1+\Sigma_{ii}^{\prime L})(1+\Sigma_{ii}^{\prime R})}
  \frac{\partial}{\partial p^2}\Sigma_{ii}^{\prime S,L}+
  \frac{m_i+\delta m_i-\Sigma_{ii}^{\prime S,L}}
  {(1+\Sigma_{ii}^{\prime L})(1+\Sigma_{ii}^{\prime R})}
  \frac{\partial}{\partial p^2}\Sigma_{ii}^{\prime S,R}]]
\ear \eeqa
If we adopt the consistent assumption: 
$\bar{Z}_{ii}^{L}\,=\,Z_{ii}^L\,,\bar{Z}_{ii}^{R}\,=\,Z_{ii}^R$, we will have

\beqa \iar
  \bar{Z}_{ii}^{L}\,=\,Z_{ii}^L\,=\,1/(A_3(1+\Sigma^{\prime L}_{ii}(m_i^2)))\,,\\
  \bar{Z}_{ii}^{R}\,=\,Z_{ii}^R\,=\,1/(A_3(1+\Sigma^{\prime R}_{ii}(m_i^2)))\,.
\ear \eeqa
with 

\beq
  A_3\,=\,1+Re[\frac{m_i^2}{1+\Sigma_{ii}^{\prime L}}\frac{\partial}{\partial p^2}
  \Sigma_{ii}^{\prime L}+\frac{m_i^2}{1+\Sigma_{ii}^{\prime R}}
  \frac{\partial}{\partial p^2}\Sigma_{ii}^{\prime R}+
  \frac{m_i+\delta m_i-\Sigma_{ii}^{\prime S,R}}
  {(1+\Sigma_{ii}^{\prime L})(1+\Sigma_{ii}^{\prime R})}
  \frac{\partial}{\partial p^2}\Sigma_{ii}^{\prime S,L}+
  \frac{m_i+\delta m_i-\Sigma_{ii}^{\prime S,L}}
  {(1+\Sigma_{ii}^{\prime L})(1+\Sigma_{ii}^{\prime R})}
  \frac{\partial}{\partial p^2}\Sigma_{ii}^{\prime S,R}]
\eeq
which are different from all of the previous results. At one-loop level, Eqs.(48)
becomes

\beqa \iar
  \bar{Z}^L_i=Z^L_i=1-\Sigma^L_{ii}(m_i^2)-m_i^2\frac{\partial}{\partial p^2}Re
  [\Sigma^L_{ii}(p^2)+\Sigma^R_{ii}(p^2)+2\Sigma^S_{ii}(p^2)]_{p^2=m_i^2}\,, \\
  \bar{Z}^R_i=Z^R_i=1-\Sigma^R_{ii}(m_i^2)-m_i^2\frac{\partial}{\partial p^2}Re
  [\Sigma^L_{ii}(p^2)+\Sigma^R_{ii}(p^2)+2\Sigma^S_{ii}(p^2)]_{p^2=m_i^2}\,.
\ear \eeqa
which are different from Eqs.(31) and (45).

Through these comparison we find that whether to introduce field renormalization
constants or not may change the fermion wrc.(comparing Eqs.(50) with Eqs.(31)). 
On the
other hand, introducing field renormalization constants will enhance the complexity
of calculating S-matrix elements. As it's very known, "bare" field isn't a physical
quantity thus doesn't need to be renormalized. So the best way to renormalize a 
quantum field theory is to discard field renormalization. It not only simplifies
the calculations of S-matrix elements, but also avoids the possible deviation 
arising from the introduction of field renormalization constants.

\section{Gauge Dependence of Renormalization Schemes}

In this section we will discuss the gauge dependence of the renormalization schemes
we have mentioned. For convenience we only discuss two typical renormalization 
schemes. The first scheme
is the scheme we have adopted, the second scheme is the scheme which adopts the
complex-pole scheme and contains field renormalization constants. At one-loop level,
the results of $\delta m_i$ and $\Gamma_i$ are the same in these two schemes. The
difference comes from the imaginary part of the fermion's wrc., if you compare
Eqs.(31) with Eq.(45). So we only need to consider the imaginary part of the 
fermion self-energy functions in the following calculations. 

We have selected two physical processes to test the gauge dependence of the 
physical amplitude in these two renormalization schemes. One process is W gauge boson 
decays into a lepton and an anti-neutrino. For our purpose it is sufficient to only 
consider the dependence of W boson gauge parameter $\xi_W$. Thus there are only two
Feynman diagrams which contribute to the imaginary part of the lepton self-energy 
functions, as shown in Fig.1 

\vspace{-2mm}
\begin{figure}[tbh]
\begin{center}
\epsfig{file=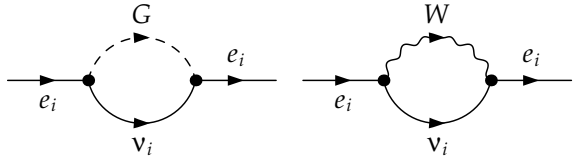,width=8cm} 
\caption{lepton's self-energy diagrams which have contributions to 
the $\xi_W$-dependent imaginary part of lepton's self-energy functions}
\end{center}
\end{figure}

Using the {\em cutting rules}\cite{c6}, we can 
calculate the imaginary part of the lepton's wrc.. In the first scheme we have:
(restrained $\xi_W>0$)

\beq
  Im[Z_i^L]\,=\,\frac{e^2 (m_{l,i}^2-M_W^2\xi_W)^2 \theta[m_{l,i}^2-M_W^2\xi_W]}
  {64\pi M_W^2 s_W^2 m_{l,i}^2}
\eeq
with $m_{l,i}$ the lepton's mass, $M_W$ the W gauge boson mass, $s_W$ the sine of 
the weak mixing angle $\theta_W$ and $\theta$ the Heaviside function.

\begin{figure}[tbh]
\begin{center}
\epsfig{file=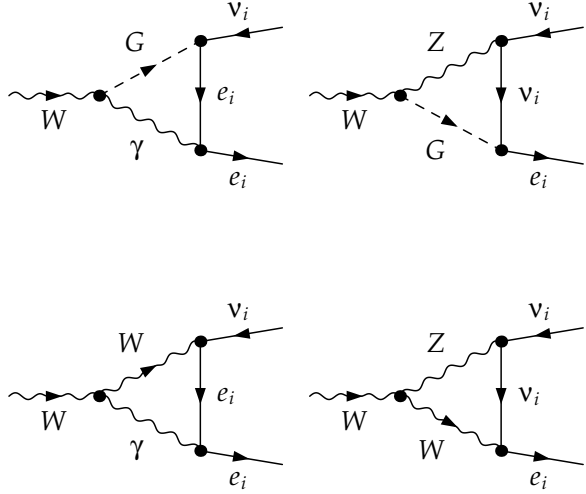,width=8cm} 
\caption{irreducible one-loop diagrams for $W^{-}\rightarrow e_i\bar{\nu}_i$
which have contributions to the $\xi_W$-dependent imaginary part of this amplitude}
\end{center}
\end{figure}

In Fig.2 we show the irreducible diagrams that contribute to the $\xi_W$-dependent 
imaginary part of
the one-loop $W^{-}\rightarrow e_i\bar{\nu}_i$ amplitude. Using the 
{\em cutting rules} we obtain 

\beq
  Im[T_1]\,=\,A_L F_L+B_L G_L+A_L Im[Z_i^L] e/(2\sqrt{2}s_W)
\eeq
with

\beqa \iar
  A_L\,=\,\bar{u}(p_2){\xslash \epsilon}\gamma_L \nu(p_1)\,,
  B_L\,=\,\bar{u}(p_2)\frac{\epsilon . p_1}{M_W}\gamma_L \nu(p_1)\,.
\ear \eeqa
and the form factors

\beqa \iar
  F_L\,=\,-\frac{e^3(\xi_W^3-3\xi_W^2+2)\theta[1-\xi_W]}{48\sqrt{2}\pi s_W}-
  \frac{e^3(m_{l,i}^2-M_W^2\xi_W)^2\theta[m_{l,i}^2-M_W^2\xi_W]}{64\sqrt{2}\pi 
  M_W^2 s_W^3 m_{l,i}^2}\,, \\
  G_L\,=\,\frac{e^3(\xi_W-1)^3 m_{l,i}M_W\theta[1-\xi_W]}{32\sqrt{2}\pi s_W
  (M_W^2-m_{l,i}^2)}\,.
\ear \eeqa
Thus we have

\beqa \iar
  Im[T_1]\,=\,A_L(-\frac{e^3(\xi_W^3-3\xi_W^2+2)\theta[1-\xi_W]}{48\sqrt{2}\pi s_W}-
  \frac{e^3(m_{l,i}^2-M_W^2\xi_W)^2\theta[m_{l,i}^2-M_W^2\xi_W]}{128\sqrt{2}\pi 
  M_W^2 s_W^3 m_{l,i}^2})+ \\ \hspace{16mm}
  B_L\frac{e^3(\xi_W-1)^3 m_{l,i}M_W\theta[1-\xi_W]}
  {32\sqrt{2}\pi s_W(M_W^2-m_{l,i}^2)}
\ear \eeqa
We find that the imaginary part of the one-loop $W^{-}\rightarrow e_i\bar{\nu}_i$ 
amplitude is gauge dependent in our scheme. On the other hand, the second scheme 
gives 

\beq
  Im[Z_{ii}^L]\,=\,\frac{e^2 (m_{l,i}^2-M_W^2\xi_W)^2 \theta[m_{l,i}^2-M_W^2\xi_W]}
  {32\pi M_W^2 s_W^2 m_{l,i}^2}
\eeq
and the imaginary part of the one-loop $W^{-}\rightarrow e_i\bar{\nu}_i$ amplitude 
(according to the above results) is

\beq
  Im[T_1^{\prime}]\,=\,-A_L\frac{e^3(\xi_W^3-3\xi_W^2+2)\theta[1-\xi_W]}
  {48\sqrt{2}\pi s_W}+B_L\frac{e^3(\xi_W-1)^3 m_{l,i}M_W\theta[1-\xi_W]}
  {32\sqrt{2}\pi s_W(M_W^2-m_{l,i}^2)}
\eeq 
also a gauge dependent quantity. At one-loop level Eq.(55) and (57) don't contribute
to the
modulus square of the physical amplitude. But at two-loop level they will have the
contributions: $Im[T_1]^{\dagger}Im[T_1]$ or 
$Im[T_1^{\prime}]^{\dagger}Im[T_1^{\prime}]$. At the limit $m_{l,i}\rightarrow 0$
they have the same contributions as follows 

\beq 
  Im[T_1]^{\dagger}Im[T_1]_{m_{l,i}\rightarrow 0}\,=\,
  Im[T_1^{\prime}]^{\dagger}Im[T_1^{\prime}]_{m_{l,i}\rightarrow 0}\,=\,
  \frac{e^6 M_W^2(\xi_W-1)^2(\xi_W^2-2\xi_W-2)^2\theta[1-\xi_W]}{2304\pi^2 s_W^2}
\eeq
Of course there will be no such gauge-dependent terms if we adopt the unitary gauge.
We hope that the contributions from the two-loop diagrams of the process 
$W^{-}\rightarrow e_i\bar{\nu}_i$ may cancel these gauge-dependent terms.

Another process we have studied is Z gauge boson decays into a pair down-type quarks.
The calculations are more complex than the previous ones. The imaginary parts of 
the down-type quark's wrc. are the same in these two schemes 
in the limit $m_{d,i}\rightarrow 0$ ($m_{d,i}$ is the down-type quark's mass). 
They are as follows 

\beqa \iar
  Im[M (Z \rightarrow d_i \bar{d}_i) ]_{m_{d,i} \rightarrow 0} \,=\, A_L[
  -\frac{1}{384\pi c_W^3 s_W^3}e^3(1-4 c_W^2\xi_W)^{3/2}\theta[1-4 c_W^2\xi_W]+
  \\ \hspace{10mm}
  \frac{1}{192\pi c_W^3 s_W}e^3((\xi_W-1)^2 c_W^4-2(\xi_W-5)c_W^2+1)
  \sqrt{(\xi_W-1)^2 c_W^4-2(\xi_W+1)c_W^2+1}\,\theta[1/c_W-\sqrt{\xi_W}-1]]
\ear \eeqa
The contribution of this quantity to the modulus square of the 
$Z\rightarrow d_i\bar{d}_i$ amplitude at two-loop level is

\beqa \iar
  Im[M(Z\rightarrow d_i\bar{d}_i)]^{\dagger}
  Im[M(Z\rightarrow d_i\bar{d}_i)]_{m_{d,i}\rightarrow 0}\,=\,
  -\frac{1}{73728\pi^2 c_W^8 s_W^6}e^6 M_W^2(4 c_W^2\xi_W-1)^3\theta[1-4 c_W^2\xi_W]
  + \\ \hspace{10mm}
  \frac{1}{18432\pi^2 c_W^8 s_W^4}e^6 M_W^2((\xi_W-1)^2 c_W^4-2(\xi_W-5)c_W^2+1)
  \theta[1/c_W-\sqrt{\xi_W}-1] \\ \hspace{10mm} 
  (s_W^2((\xi_W-1)^4 c_W^8-4(\xi_W-2)(\xi_W-1)^2 c_W^6+(6\xi_W^2-20\xi_W-18) c_W^4-
  4(\xi_W-2) c_W^2+1)- \\ \hspace{10mm} 
  (1-4 c_W^2\xi_W)^{3/2}\sqrt{(\xi_W-1)^2 c_W^4-2(\xi_W+1) c_W^2+1})
\ear \eeqa
There are also gauge-dependent terms. But we don't know how they can 
be cancelled in the full modulus square of the $Z\rightarrow d_i\bar{d}_i$ 
amplitude at two-loop level. This problem will be left for our next work.

Through these discussions we can draw a conclusion that based on the present 
knowledge we cannot judge which scheme is the better.

\section{Introduction of Off-diagonal Wave-function Renormalization Constants}

In order to simplify the calculations of the corrections of fermion external legs 
in a 
Feynman diagram we can change the fermion wrc. to two matrices, which will contain
the contributions of the off-diagonal two-point diagrams at fermion external legs. 
In order to
simplify this definition we need to define two new self-energy functions as follows:

\vspace{4mm} 
\beqa \iar
  \begin{picture}(420,10)
    \Text(10,5)[]{$i\Sigma_{ij}\,=\,$}
    \ArrowLine(25,5)(38,5)
    \GCirc(42,5){4}{0.5}
    \ArrowLine(46,5)(59,5)
    \Text(30,13)[]{$j$}
    \Text(54,13)[]{$i$}
    \Text(67,5)[]{$+$}
    \Text(80,1)[]{$\stackrel{\sum}{k}$}
    \ArrowLine(85,5)(98,5)
    \GCirc(102,5){4}{0.5}
    \ArrowLine(106,5)(119,5)
    \GCirc(123,5){4}{0.5}
    \ArrowLine(127,5)(140,5)
    \Text(90,13)[]{$j$}
    \Text(112,13)[]{$k$}
    \Text(135,13)[]{$i$}
    \Text(148,5)[]{$+$}
    \Text(161,1)[]{$\stackrel{\sum}{k,l}$}
    \ArrowLine(166,5)(179,5)
    \GCirc(183,5){4}{0.5}
    \ArrowLine(187,5)(200,5)
    \GCirc(204,5){4}{0.5}
    \ArrowLine(208,5)(221,5)
    \GCirc(225,5){4}{0.5}
    \ArrowLine(229,5)(242,5)
    \Text(171,13)[]{$j$}
    \Text(193,13)[]{$k$}
    \Text(214,13)[]{$l$}
    \Text(237,13)[]{$i$}
    \Text(250,5)[]{$+$}
    \Text(266,5)[]{$\cdot \cdot \cdot$}
    \Text(282,5)[]{$+$}
    \Text(298,1)[]{$\stackrel{\sum}{k,l,\cdot \cdot \cdot,m}$}
    \ArrowLine(306,5)(319,5)
    \GCirc(323,5){4}{0.5}
    \ArrowLine(327,5)(340,5)
    \GCirc(344,5){4}{0.5}
    \ArrowLine(348,5)(361,5)
    \Text(371,5)[]{$\cdot \cdot \cdot$}
    \ArrowLine(377,5)(390,5)
    \GCirc(395,5){4}{0.5}
    \ArrowLine(399,5)(412,5)
    \Text(311,13)[]{$j$}
    \Text(333,13)[]{$k$}
    \Text(354,13)[]{$l$}
    \Text(383,13)[]{$m$}
    \Text(407,13)[]{$i$}
    \Text(380,-15)[]{$k,l,\cdot \cdot \cdot,m \not= j$}
  \end{picture} \\ \vspace{6mm} \\
  \begin{picture}(420,10)
    \Text(10,5)[]{$i\bar{\Sigma}_{ij}\,=\,$}
    \ArrowLine(25,5)(38,5)
    \GCirc(42,5){4}{0.5}
    \ArrowLine(46,5)(59,5)
    \Text(30,13)[]{$j$}
    \Text(54,13)[]{$i$}
    \Text(67,5)[]{$+$}
    \Text(80,1)[]{$\stackrel{\sum}{k}$}
    \ArrowLine(85,5)(98,5)
    \GCirc(102,5){4}{0.5}
    \ArrowLine(106,5)(119,5)
    \GCirc(123,5){4}{0.5}
    \ArrowLine(127,5)(140,5)
    \Text(90,13)[]{$j$}
    \Text(112,13)[]{$k$}
    \Text(135,13)[]{$i$}
    \Text(148,5)[]{$+$}
    \Text(161,1)[]{$\stackrel{\sum}{k,l}$}
    \ArrowLine(166,5)(179,5)
    \GCirc(183,5){4}{0.5}
    \ArrowLine(187,5)(200,5)
    \GCirc(204,5){4}{0.5}
    \ArrowLine(208,5)(221,5)
    \GCirc(225,5){4}{0.5}
    \ArrowLine(229,5)(242,5)
    \Text(171,13)[]{$j$}
    \Text(193,13)[]{$k$}
    \Text(214,13)[]{$l$}
    \Text(237,13)[]{$i$}
    \Text(250,5)[]{$+$}
    \Text(266,5)[]{$\cdot \cdot \cdot$}
    \Text(282,5)[]{$+$}
    \Text(298,1)[]{$\stackrel{\sum}{k,l,\cdot \cdot \cdot,m}$}
    \ArrowLine(306,5)(319,5)
    \GCirc(323,5){4}{0.5}
    \ArrowLine(327,5)(340,5)
    \GCirc(344,5){4}{0.5}
    \ArrowLine(348,5)(361,5)
    \Text(371,5)[]{$\cdot \cdot \cdot$}
    \ArrowLine(377,5)(390,5)
    \GCirc(395,5){4}{0.5}
    \ArrowLine(399,5)(412,5)
    \Text(311,13)[]{$j$}
    \Text(333,13)[]{$k$}
    \Text(354,13)[]{$l$}
    \Text(383,13)[]{$m$}
    \Text(407,13)[]{$i$}
    \Text(380,-15)[]{$k,l,\cdot \cdot \cdot,m \not= i$}
  \end{picture} 
\ear \eeqa

\vspace{6mm} 

\hspace{-4mm}and the corresponding renormalized self-energy functions (after adding 
proper counterterms to the above diagrams):

\beqa \iar
  \hat{\Sigma}_{ij}\,=\, \xslash p (\Sigma^{L}_{ij}\gamma_L + 
  \Sigma^{R}_{ij}\gamma_R) + \Sigma^{S,L}_{ij}\gamma_L + 
  \Sigma^{S,R}_{ij}\gamma_R \,, \, i\not=j\,, \\
  \hat{\bar{\Sigma}}_{ij}\,=\, \xslash p (\bar{\Sigma}^{L}_{ij}\gamma_L + 
  \bar{\Sigma}^{R}_{ij}\gamma_R) + \bar{\Sigma}^{S,L}_{ij}\gamma_L + 
  \bar{\Sigma}^{S,R}_{ij}\gamma_R \,, \, i\not=j\,.
\ear \eeqa
We will see in a second that the new self-energy functions have no other meanings 
except for bringing about great convenience in actual calculations.

Now we introduce the fermion wrc. matrices $\bar{Z}_{ij}^{\half}$ and 
$Z_{ij}^{\half}$ as follows:

\beqa \iar
  <p_i ... |S|k_j ...>\,=\, \bar{u}(m_i)\bar{Z}_i^{\half}
  M_{ij}Z_j^{\half}u(m_j) \\
  \hspace{24mm} \,=\, \bar{u}(m_i)\bar{Z}_i^{\half}
  (\delta_{ii^{\prime}} + i\hat{\bar{\Sigma}}_{ii^{\prime}}(\xslash p_i)
  \frac{i\delta_{i \not= i^{\prime}}}{\xslash p_i - m_{i^{\prime}}}) 
  M_{i^{\prime}j^{\prime}}^{amp}
  (\delta_{j^{\prime}j} + \frac{i\delta_{j^{\prime}\not=j}}{\xslash k_j - 
  m_{j^{\prime}}}
  i\hat{\Sigma}_{j^{\prime}j}(\xslash k_j)) Z_j^{\half}u(m_j) \\
  \hspace{24mm}\,\equiv \, \bar{u}(m_i)\bar{Z}_{ii^{\prime}}^{\half}
  M_{i^{\prime}j^{\prime}}^{amp}Z_{j^{\prime}j}^{\half}u(m_j)\,,
\ear \eeqa
with $\delta_{i\not=j}=1$ when $i\not=j$ and $M^{amp}$ 
the amplitude of completely amputated diagrams (without off-diagonal self-energy 
diagrams at fermion external legs). Noted we have plainly used the definitions of 
Eqs.(62) in the second line of Eqs.(63). The last line of 
Eqs.(63) gives the definitions of fermion wrc. matrices:

\beqa \iar
  Z_{ij}^{\half}u(m_j)\,=\,(\delta_{ij} + \frac{i\delta_{i\not=j}}{\xslash p_j - 
  m_{i}}
  i\hat{\Sigma}_{ij}(\xslash p_j))Z_j^{\half}u(m_j)|_{p_j^2=m_j^2} \,, \\
  \bar{u}(m_i)\bar{Z}_{ij}^{\half}\,=\, \bar{u}(m_i)\bar{Z}_i^{\half}
  (\delta_{ij} + i\hat{\bar{\Sigma}}_{ij}(\xslash p_i)
  \frac{i\delta_{i\not= j}}{\xslash p_i - m_j})|_{p_i^2=m_i^2} \,.
\ear \eeqa

after some algebra we have

\beqa \iar
  \bar{Z}_{ij}^{L\half} = Z_i^{L\half}\delta_{ij}+\frac{\delta_{i\not=j}}{m_j^2- 
  m_i^2}[m_i^2 Z_i^{L\half}\bar{\Sigma}_{ij}^L(m_i^2) + 
  m_i m_j Z_i^{R\half}\bar{\Sigma}_{ij}^R(m_i^2) + 
  m_i Z_i^{R\half}\bar{\Sigma}_{ij}^{S,L}(m_i^2) + 
  m_j Z_i^{L\half}\bar{\Sigma}_{ij}^{S,R}(m_i^2)]\,, \\ 
  \bar{Z}_{ij}^{R\half} = Z_i^{R\half}\delta_{ij}+\frac{\delta_{i\not=j}}{m_j^2- 
  m_i^2}[m_i m_j Z_i^{L\half}\bar{\Sigma}_{ij}^L(m_i^2) + 
  m_i^2 Z_i^{R\half}\bar{\Sigma}_{ij}^R(m_i^2) + 
  m_j Z_i^{R\half}\bar{\Sigma}_{ij}^{S,L}(m_i^2) + 
  m_i Z_i^{L\half}\bar{\Sigma}_{ij}^{S,R}(m_i^2)]\,,
\ear \eeqa
and

\beqa \iar
  Z_{ij}^{L\half} = Z_j^{L\half}\delta_{ij} + \frac{\delta_{i\not=j}}{m_i^2 - m_j^2}
  [m_j^2 Z_j^{L\half}\Sigma_{ij}^L(m_j^2)+m_i m_j Z_j^{R\half}\Sigma_{ij}^R(m_j^2)+
  m_i Z_j^{L\half}\Sigma_{ij}^{S,L}(m_j^2)+m_j Z_j^{R\half}\Sigma_{ij}^{S,R}(m_j^2)]
  \,, \\ 
  Z_{ij}^{R\half} = Z_j^{R\half}\delta_{ij} + \frac{\delta_{i\not=j}}{m_i^2 - m_j^2}
  [m_i m_j Z_j^{L\half}\Sigma_{ij}^L(m_j^2)+m_j^2 Z_j^{R\half}\Sigma_{ij}^R(m_j^2)+
  m_j Z_j^{L\half}\Sigma_{ij}^{S,L}(m_j^2)+
  m_i Z_j^{R\half}\Sigma_{ij}^{S,R}(m_j^2)]\,.
\ear \eeqa
At one-loop level these results agree with Eqs.(3.3) and Eqs.(3.4) in Ref.\cite{c3}.

Using these wrc. matrices we can calculate S-matrix 
elements without considering any of the corrections of fermion external legs
in a Feynman diagram. Noted such matrices are also suitable for gauge bosons
which have mixing between each other. 

\section{Conclusion}

In this paper we have attempted to extend LSZ reduction formula for unstable fermion.
We adopted the assumption of Eq.(7), which resembles the relativistic Breit-Wigner
formula thus can describe the behavior of unstable particle's propagator in the 
resonant region. Under this assumption only the on-shell mass renormalization scheme
is suitable (see section 2). In consideration of the difficulty of introducing 
fermion field renormalization constants\cite{c3}
we have abandoned field renormalization. We assume the unstable
fermion's propagator has the form of Eq.(9) in the limit $p^2\rightarrow m_i^2$.
Thus we can get the LSZ reduction formula Eq.(12) for unstable fermion and obtain 
the proper fermion wrc. Eqs.(29). 

Next we change our assumption and compare the difference between different 
renormalization schemes. There are four basic renormalization schemes, corresponding
to whether or not to introduce field renormalization constants and to adopt which 
mass renormalization scheme, on-shell scheme or complex-pole scheme\cite{c2}. Through
comparison we find that whether to introduce field renormalization constants or not
will affect the values of fermion wrc. and demonstrate its difference in the 
imaginary part of physical amplitude, and so does the choice of mass
renormalization scheme (see section 3 and 4). This discrepancy is gauge dependent 
and can manifest 
itself at the modulus square of physical amplitude at two-loop level. 
But we haven't found strong evidence to judge which scheme is better that 
the others.

Next we change the fermion wrc. to two matrices, in order to contain the corrections
of fermion external legs in a Feynman diagram. The new fermion wrc. matrices,
according to Eqs.(63), will simplify the calculations of physical amplitude, because
we don't need to calculate the corrections of off-diagonal self-energy diagrams
at fermion external legs. We note that the concept of wrc. Matrix can be extended
to bosons provided they have mixing in themselves.

In conclusion, we recommend the renormalization scheme which adopts the on-shell
mass renormalization scheme and abandons field renormalization, because this 
renormalization scheme is very simple and clear. We will study this point further 
in the future.

\vspace{6mm}
{\bf \Large Acknowledgments}
\vspace{2mm}

The author thanks professor Xiao-Yuan Li for his useful guidance 
and Dr. Xiao-Hong Wu for his sincerely help.


\end{document}